\documentclass[conference]{ieeetran}  
\IEEEoverridecommandlockouts
\usepackage{cite}
\usepackage{amsmath,amssymb,amsfonts}
\usepackage{algorithmic}
\usepackage[ruled,vlined,linesnumbered]{algorithm2e}
\usepackage{graphicx}
\usepackage{textcomp}
\usepackage{xcolor}
\usepackage[UKenglish,english]{babel}
\usepackage{amsmath,amssymb,paralist,subfigure,graphicx,stmaryrd,mathrsfs,url,algorithm2e,soul}
\usepackage{flushend}
\usepackage{graphicx}
\usepackage{hyperref} 
\usepackage{cuted,mathtools,lipsum,cite}

\newtheorem{rem}{Remark}

\def\mb{\mathbf}

\def\mc{\mathcal}

\begin{document}
\title{\LARGE \bf Linear TDOA-based Measurements for Distributed Estimation and Localized Tracking   } 

\author{\IEEEauthorblockN{Mohammadreza Doostmohammadian$^{1,2}$, Themistoklis Charalambous$^{1,3}$}
\IEEEauthorblockA{$^{1}$\textit{School of Electrical Engineering, Aalto University, Espoo, Finland, 
\texttt{name.surname@aalto.fi}}
\\ $^{2}$\textit{Mechatronics Department, Faculty of Mechanical Engineering, Semnan University, Iran, \texttt{doost@semnan.ac.ir}}
\\ $^{3}$\textit{Department of Electrical and Computer Engineering, University of Cyprus, Cyprus \texttt{surname.name@ucy.ac.cy}}
} 
\thanks{This work is supported in part by
the European Commission through the H2020 Project Finest Twins under
Agreement 856602.}
}

\maketitle
\thispagestyle{empty}
\begin{abstract}
	We propose a linear time-difference-of-arrival (TDOA)  measurement model to improve \textit{distributed} estimation performance for localized target tracking. We design distributed filters over sparse (possibly large-scale) communication networks using consensus-based data-fusion techniques. The proposed distributed and localized tracking protocols considerably reduce the sensor network's required connectivity and communication rate. We, further, consider  $\kappa$-redundant observability and fault-tolerant design in case of losing communication links or sensor nodes. We present the minimal conditions on the remaining sensor network (after link/node removal) such that the distributed observability is still preserved and, thus, the sensor network can track the (single) maneuvering target. The motivation is to reduce the communication load versus the processing load, as the computational units are, in general, less costly than the communication devices. We evaluate the tracking performance via simulations in MATLAB.
\end{abstract}

\section{Introduction} \label{sec_intro}
\IEEEPARstart{L}{ocalized} tracking and distributed estimation is considered in this paper to track a moving target via some static sensors, for example, tracking a rocket, a drone/UAV, or an airplane. Distributed estimators (or filters) has recently gained significant interest in signal processing, control, and machine learning literature to avoid single-node-of-failure \cite{spl17} and localize the computation and data processing \cite{jiang2022multi,safavi2018distributed,ennasr2020time,ennasr2016distributed} based on consensus algorithms \cite{taes20}.
This is further motivated by the recent advancement in parallel processing, cloud computing, the Internet of Things (IoT), and wireless communication networks. The existing distributed methods are mainly based on consensus data-fusion and are categorized as either single-time-scale (STS) or double-time-scale (DTS) methods, based on their communication/consensus time-scale with respect to the target dynamics. The STS methods \cite{ennasr2020time,ennasr2016distributed,mo2020,mohammadi2015distributed} perform one step of consensus at every time step of system (target) dynamics (the same time-scale), while the DTS methods \cite{usman_acc:11,he2020secure,olfati2002distributed,Battilotti} perform many iterations of consensus fusion between two consecutive time-steps of the dynamics, see the differences in \cite[Fig. 1]{TCNS2020}. These distributed strategies are mainly developed over linear and partially observable (measurement/system) models, e.g., for model-based and kinematic-based vehicle speed estimation \cite{hashemi2017distributed}. In this direction, the nonlinear range-based methods as TDOA \cite{ennasr2020time,ennasr2016distributed,5778025}, are \textit{linearized} to be adapted for distributed setups. Such a linearization, in general, degrades the performance of the estimation while resulting in a time-varying output matrix and, in turn, a time-varying feedback gain. This requires redesigning the gain matrix at every iteration, e.g., via the iterative linear-matrix-inequalities (LMI) design \cite{jstsp,jstsp14}, and imposes more computational loads on the sensors. For practical target tracking applications, we need networked estimation scenarios with less communication and computational load on sensors, both in terms of communication traffic (networking) and data exchange rate.
This plays a key role in real-time tracking of long-range maneuvering targets; see a survey in \cite{survey_vtm}. Another challenge is to develop fault detection and isolation (FDI) algorithms for the distributed estimation networks, i.e., a \textit{joint} distributed estimation and \textit{localized} FDI setup \cite{icas21_attack,anomaly_ECC22}. A survey of security measures for the centralized observer design, with applications to intelligent transportation systems and tracking, is discussed in \cite{dibaji2019systems,giraldo2018survey}.  

\textbf{Main contributions:} We propose a new \textit{linear} TDOA-based measurement model, in contrast to existing \textit{nonlinear} counterparts. Recall that such nonlinear models, although acceptable in the centralized setup, need to be \textit{linearized} for most distributed protocols. We compare the performance of the proposed linear model for modified TDOA with the existing nonlinear one, e.g., in \cite{ennasr2020time}. Linearization of such models may significantly reduce the performance of distributed tracking due to inaccurate measurements. Another reason for the inaccuracy is that the \textit{output matrix} in the linearized model is also a function of the target position \textit{to be estimated}. The proposed distributed estimator in this paper only requires \textit{distributed observability} in contrast to \textit{local observability} in the existing STS \cite{ennasr2020time,ennasr2016distributed,mo2020,mohammadi2015distributed} and MTS \cite{usman_acc:11,he2020secure,olfati2002distributed,Battilotti} methods. This reduces (i) required communication traffic and linking over the network for the STS scenario and (ii) the required communication and consensus rate for the MTS models. 
%We compare this improved performance in terms of connectivity and computation. 
The proposed method, further, can be improved by designing $\kappa$-redundant distributed observers, which are resilient to the removal (or failure) of up-to $\kappa$ nodes (sensors) or links (communications) over the network. Moreover, \textit{localized} distributed fault detection can be applied to this model to isolate possible faulty measurements over the sensor network. Thus, the proposed STS estimation and fault detection methodology are \textit{localized} and do not need any centralized processing or decision-making unit, which is of interest in large-scale applications over faulty environments.

\textbf{Organization:} The setup for the target tracking is discussed in Section~\ref{sec_fram}. The proposed distributed estimation along with the fault-detection setup are discussed in Sections~\ref{sec_dist} and \ref{sec_fault}. Sections \eqref{sec_sim} and \eqref{sec_con} provide the simulations and conclusions.

\section{The Framework} \label{sec_fram}
We model the target based on the nearly-constant-velocity (NCV) model. This   formulation is widely used in the literature, e.g., see \cite{ennasr2016distributed,ennasr2020time,gustafsson2002particle,bar2004estimation}, representing a linear integrator model as,
\begin{equation}  \label{eq_targ}
		\mb{x}(k+1) = F\mb{x}(k)+G\mb{w}(k)
\end{equation}
with state-vector $\mb{x} = \left(p_x;p_y;p_z;\dot{p}_x;\dot{p}_y;\dot{p}_z\right)$ as the position and velocity of the target in 3D space at time $k$ (``;'' as the column concatenation), $\mb{w}(k)=\mc{N}(0,Q)$ as some random input, and $F$ and $G$ respectively representing the transition and the input matrix defined as \cite{gustafsson2002particle},
\begin{align}  \label{F_ncv}
		F = \left(
		\begin{array}{cc}
			\mb{I}_3 & T \mb{I}_3\\
			\mb{0}_3 & \mb{I}_3 \\ 	
		\end{array} \right),
		G = \left(
		\begin{array}{c}
			\frac{T^2}{2} \mb{I}_3 \\
			T\mb{I}_3 \\ 	
		\end{array} \right)
\end{align}
with $T$ as the sampling period, and $I_n$ and $\mb{0}_n$ as the identity and $0$ matrix of size $n$. The model can be simply extended to the nearly-constant-acceleration (NCA) model or any other target dynamics model (see \cite{bar2004estimation} for more examples). 

\begin{figure} [t]
		\centering
		\includegraphics[width=3.4in]{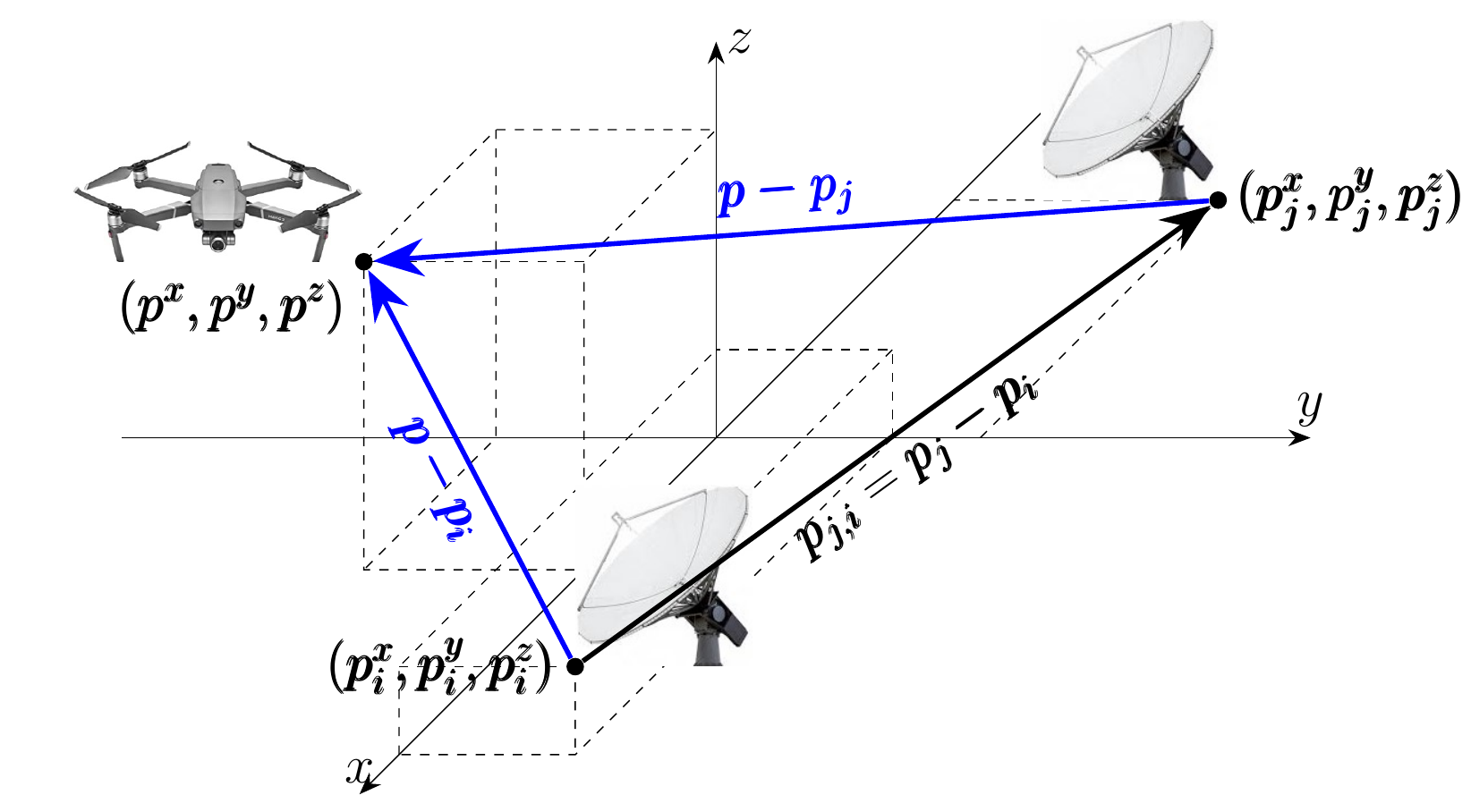}
		\caption{This figure shows a group of static sensors (radars) receiving a beacon signal from the target (drone). The time-difference-of-arrival (TDOA) measurements are based on the positions of the target $\mb{p}$ and the neighboring sensors $\mb{p}_i,\mb{p}_j$. $\mb{p}_{j,i}$ represents the (constant) relative position of the sensor $j$ with respect to the sensor $i$. }
		\label{fig_radar_drone}
\end{figure}

We consider a network of $N$ sensors with position of sensor $i$ denoted by $\mb{p}_i = (p_{x,i};p_{y,i};p_{z,i})$ where the sensors are not located on the same $x,y,z$ plane. Each sensor receives a beacon signal with \textit{known propagation speed} $c$ from the mobile target at every time $k$ and computes the time-of-arrival (TOA) measurement (or range) based on $\|\mb{p}(k)-\mb{p}_i(k)\| = ct_i$ ($\|\cdot \|$ as the $2$-norm) and shares this information, along with its position information, with the set of its direct neighboring sensors $\mc{N}_i$. Overall, each sensor possess a list of TOAs and positions of the neighboring sensors. in the centralized setup the sensors share their information with many other sensors (e.g., an all-to-all network). In the existing literature, each sensor $i$ performs filtering based on its nonlinear TDOA measurement vectors as,
\begin{align} \label{eq_y_nonlin}
		\mb{y}_i(k) &= h_i(k) + \nu_i(k) 
		 \\
		 h_i(k) &= (h_{i,j_1}(\mb{x}(k));\dots;h_{i,j_{|\mc{N}_i|}}(\mb{x}(k))) + \nu_i(k)
		 \\ \label{eq_tdoa}
		h_{i,j}(\mb{x}(k)) &= \|\mb{p}(k)-\mb{p}_i(k)\|-\|\mb{p}(k)-\mb{p}_{j}(k)\|.
\end{align}
i.e., subtracting the TOAs of its neighbors  $\{j_1,\dots,j_{|\mc{N}_i|}\}$ from its own TOA (with additive Gaussian measurement noise $\nu_i(k)=\mc{N}(0,R)$). See Fig.~\ref{fig_radar_drone} for more illustrations.  
In this paper, we first improve the proposed nonlinear measurement model \eqref{eq_y_nonlin}-\eqref{eq_tdoa}, and provide a linear TDOA measurement model to be used in distributed estimation and filtering. Then we propose a distributed estimator (along with localized fault detection strategy) to track the target via a sensor network with no need for local observability of the target in the neighborhood of any sensor (this is referred to as \textit{distributed observability} \cite{globalsip14}). Further, $\kappa$-redundant estimator design and local fault-detection and isolation (FDI) are also considered. 
 
\section{The Proposed Distributed Setup} \label{sec_dist}
For existing linear distributed estimation methods, the nonlinear measurement model \eqref{eq_y_nonlin}-\eqref{eq_tdoa} needs to be linearized as,
	\begin{align} \label{eq_y_lin}
		\mb{y}_i(k) &= H_i(k)\mb{x}(k) + \nu_i(k),\\ \label{eq_h_ncv}
		H_i(k) &=\left(
		\begin{array}{cccccc}
			\frac{\partial h_{i,j_1}}{\partial p^{x}} & \frac{\partial h_{i,j_1}}{\partial p^{y}} & \frac{\partial h_{i,j_1}}{\partial p^{z}} &0& 0& 0\\
			\vdots & \vdots & \vdots & \vdots & \vdots & \vdots\\
			\frac{\partial h_{i,j_{|\mc{N}_i|}}}{\partial p^{x}} & \frac{\partial h_{i,j_{|\mc{N}_i|}}}{\partial p^{y}} & \frac{\partial h_{i,j_{|\mc{N}_i|}}}{\partial p^{z}} &0& 0& 0
		\end{array} \right),
	\end{align}
where,	
	\begin{align}
		\frac{\partial h_{i,j}}{\partial p_{x}} = \frac{p_x-p_{x,i}}{\|\mb{p}-\mb{p}_i\|}-\frac{p_x-p_{x,j}}{\|\mb{p}-\mb{p}_{j}\|},\\
		\frac{\partial h_{i,j}}{\partial p_{y}} = \frac{p_y-p_{y,i}}{\|\mb{p}-\mb{p}_i\|}-\frac{p_y-p_{y,j}}{\|\mb{p}-\mb{p}_{j}\|},\\ \label{eq_hij}
		\frac{\partial h_{i,j}}{\partial p_{z}} = \frac{p_z-p_{z,i}}{\|\mb{p}-\mb{p}_i\|}-\frac{p_z-p_{z,j}}{\|\mb{p}-\mb{p}_{j}\|}.
	\end{align}
Note that all the parameters of the above measurement (output) matrix $H_i$ depend on the target position $\mb{p}$. Recall that, since the exact position of the target $\mb{p}$ needs to be estimated by sensors (and is unknown), instead the estimated position $\widehat{\mb{p}}_i$ by each sensor $i$ is used in \eqref{eq_tdoa} for the feedback gain design in \cite{ennasr2016distributed,ennasr2020time}. This, in turn, adds more uncertainty and accentuates the error of the linearized model. We dropped the dependence on time $k$ only for notation simplicity. To provide a more accurate \textit{linear} model for distributed estimation methods and localized tracking applications we propose the following (instead of \eqref{eq_tdoa}),
	
	\small \begin{align} \label{eq_h_simp}
		h_{i,j}(\mb{x}(k)) =& \frac{1}{2}\Big(\|\mb{p}(k)-\mb{p}_i(k)\|^2-\|\mb{p}(k)-\mb{p}_{j}(k)\|^2\Big)  \\ \nonumber
		=& \frac{1}{2} \Big((p_{x,j}-p_{x,i})(2p_x-p_{x,j}-p_{x,i}) \\ \nonumber
       & ~ +(p_{y,j}-p_{y,i})(2p_y-p_{y,j}-p_{y,i}) \\ 
       & ~ + (p_{z,j}-p_{z,i})(2p_z-p_{z,j}-p_{z,i}) \Big)\\
       =& H_i \mb{p}(k) - \frac{1}{2}(\mb{p}_j^2 - \mb{p}_i^2) 
       \label{eq_h_simp2}
	\end{align}	\normalsize
	where\footnote{There is a  biasing term $- \frac{1}{2}(\mb{p}_j^2 - \mb{p}_i^2)$ that is \textit{known} (based on the shared position of neighboring sensors) and can be easily subtracted from the TDOA measurement \eqref{eq_h_simp}. Then, we get the standard linear output form $\mb{y}_i(k) = H_i\mb{x}(k) + \nu_i(k)$.},
	\begin{align} \label{eq_tdoa_new}
		H_i=\left(
		\begin{array}{cccccc}
			 p_{x,j,i} & p_{y,j,i} &  p_{z,j,i} & 0 & 0 & 0\\
			\vdots & \vdots & \vdots & \vdots & \vdots & \vdots\\
			 p_{x,j_{|\mc{N}_i|},i} & p_{y,j_{|\mc{N}_i|},i} &  p_{z,j_{|\mc{N}_i|},i} & 0 & 0 & 0\\
		\end{array} \right),
	\end{align}	
Note that the above measurement model is linear with constant $H_i$ for static sensors, which is an improvement over the linearized time-varying $H_i$ in \eqref{eq_h_ncv}-\eqref{eq_hij}. Note that, from Fig.~\ref{fig_radar_drone}, $\mb{p}_{j,i}$ is fixed for two neighboring sensors $j,i$. We specifically use this for our distributed estimation setup in the following. 

For the linear target model \eqref{eq_targ}-\eqref{F_ncv} and measurements \eqref{eq_h_simp}-\eqref{eq_tdoa_new} we propose an STS networked estimator with consensus-fusion on prior estimates and innovation-update based on the measurement matrix $H_i$ in \eqref{eq_h_simp}-\eqref{eq_tdoa_new} to get the posterior estimate. the proposed localized protocol is,
\begin{align}\label{eq_p}
		\widehat{\mb{x}}_i(k|k-1) &= \sum_{j\in\mathcal{N}_i} W_{ij}F\widehat{\mb{x}}_j(k-1), ~j\in \mc{N}_i\\ \label{eq_m}
		\widehat{\mb{x}}_i(k) &= \widehat{\mb{x}}_i(k|k-1) + K_{i} H_i^\top \left(\mb{y}_i(k)-H_i\widehat{\mb{x}}_i(k|k-1)\right),
\end{align}
with $W_{ij}$ matrix representing the adjacency  matrix of the communication network of sensors $\mc{G}$ (or the sensor network) and local gain matrix $K_i$ to be designed later. Recall that for consensus-update the $W$ matrix needs to be further \textit{row-stochastic} \cite{jstsp,jstsp14}. $\widehat{\mb{x}}_i(k)$ gives the (posterior) estimate of sensor $i$ on the target state (position and speed), i.e., $\widehat{\mb{x}}_i(k)=(\widehat{\mb{p}}_i;\dot{\widehat{\mb{p}}}_i)$. It can be shown that the (collective) error dynamics for this estimator is in the form,  
	\begin{align}\label{eq_err1}
		\mb{e}(k) = (W\otimes F  - K D_H(W\otimes F))\mb{e}(k-1)+
		\mb{\eta}(k),
	\end{align}
with block-diagonal matrices $D_H := \mbox{diag}[H_i^\top H_i]$, $K := \mbox{diag}[K_{i}]$ representing the local feedback gains,	and $\mb{\eta}$ collecting all the noise-related terms defined as,

\small \begin{align}
		\mb{\eta}(k) &= \mb{1}_n \otimes G\mb{w}(k-1) - K\big(D_H(\mb{1}_n \otimes G\mb{w}(k-1)) + \overline{D}_H\nu(k)\big) \label{eq_eta}
	\end{align} \normalsize
where $\nu=(\nu_{1}; \dots;\nu_{n})$, $\mb{1}_n$ is the column-vector of ones of size $n$, and
$\overline{D}_H \triangleq \mbox{diag}[ H_i]$ is defined as the block-diagonal matrix of all measurement matrices $H_i$. Note that $\mb{e}(k)$ and $\mb{\eta}(k)$ represent the column concatenation of the local error terms as $\mb{e}(k)=(\mb{e}_{1}(k); \dots;\mb{e}_{n}(k))$ and $\mb{\eta}(k)=(\mb{\eta}_{1}(k); \dots;\mb{\eta}_{n}(k))$. From the Kalman stability theorem \cite{bay}, error dynamics \eqref{eq_err1} is stabilizable if the pair $(W\otimes F,D_H)$ is observable. \textit{This is known as distributed observability and implies observability over a network (associated with $W$). This is a much milder condition than the local observability needed in many STS distributed estimators, e.g., \cite{mohammadi2015distributed}. }
Following the structured systems theory, the error stability can be defined irrespective of the exact numeric of system entries and based on the \textit{zero-nonzero structure} of matrices $W$, $F$, and $H_i$. Such a structural (or generic design) results in stable error dynamics for almost all values of system parameters (i.e., the entries of the mentioned matrices) \cite{pequito2014optimal,tsipn19}. Furthermore, following structural analysis on the Kronecker network products given by \cite{tsipn19}, it can be shown that $(W\otimes F,D_H)$ is observable if matrix $W$ is irreducible (distributed observability), which implies that the network (of sensors) needs to be strongly-connected (SC). This topology means that the information of every sensor reaches other sensors over the network via a \textit{path of sensor nodes instead of direct information exchange} in literature. On the other hand, DTS methods require $L\gg 1$ iterations of communication and consensus between time-steps $k$ and $k+1$, see a comparison in Table~\ref{tab_compare}. 
\begin{table} [t] 
		\centering
		\caption{Comparison between different distributed tracking methods: possibility of joint FDI, connectivity $\times$ communication-rate. }
		\label{tab_compare}
		\begin{tabular}{|c|c|c|} 
			\hline
			 time-scale & filtering $+$ detection & communication links $\times$ rate    \\
			\hline
			STS & - &  $3(N-1)\times 1$   \\
			\hline
			STS (this work) & $\checkmark$ &  $N\times 1$  \\\hline
			DTS & $\checkmark$ &  $N\times L$ with $L\gg 1$\\
			\hline
			\hline
		\end{tabular}
\end{table}
This imposes high communication and processing load on sensors. 
In this work, an SC network topology ensures the existence of the local gain matrix $K$ for Schur stability of the tracking error dynamics~\eqref{eq_err1}, where such (block-diagonal) gain $K$ can be designed via the iterative LMIs in \cite[Appendix]{jstsp}, either offline (to be embedded at sensors before the tracking procedure) or online (along with the same iterations $k$ of the target dynamics). Note that one can improve the convergence rate by adding constraints to the LMI gain design subject to bound on the spectral radius of the closed-loop error dynamics, i.e., $\rho(W\otimes F  - K D_H(W\otimes F))<\overline{\rho}<1$ \cite{LCSS21}, which can further handle communication time-delays. Furthermore, the proposed distributed method allows for simultaneous detection, fault-tolerant design, and estimation; see the next section.  
\begin{rem} \label{rem_compare} 
    The LMI design of the gain matrix $K$ is based on the matrices $F$ and $H_i$. For the linearized model  \eqref{eq_h_ncv}-\eqref{eq_hij}, $H_i$ is time-varying (as it depends on the target position $\mb{p}$). This time-dependency makes the LMI design computationally burdensome since it needs to be repeated at every time $k$ for the updated (time-dependent) measurement matrix $H_i$. In our linear model \eqref{eq_h_simp}-\eqref{eq_tdoa_new}, the measurement matrix $H_i$ is time-invariant and is designed only once; thus, significantly reducing the computational load on the sensors (e.g., for real-time applications) as compared to \cite{ennasr2016distributed,ennasr2020time}. Further, since the target location is not known (and needs to be estimated), the sensors use their estimates $\widehat{\mb{p}}_i$ (the position-related elements of $\widehat{\mb{x}}_i$) for the gain design, which further increases the tracking error; see more illustrations by simulation in Section~\ref{sec_sim}. 
\end{rem}    
\section{Fault-Tolerant Design} \label{sec_fault}
    The next step is to locally detect and isolate possible measurement faults over the sensor network. In this case, consider \textit{potential} additive faults at measurements as,
    \begin{align} \label{eq_y_lin_fault}
		\mb{y}_i(k) = H_i(k)\mb{x}(k) + \nu_i(k) + f_i(k),
    \end{align}
    with $f_i(k)$ denoting the fault-term. Then, following the distributed FDI strategy in both \textit{stateless} \cite{tnse21} and \textit{stateful} setups \cite{icas21_attack}, each sensor $i$ can find possible faults at its measurements by  monitoring its (absolute) instantaneous residual $r_i(k) = |y_i(k)-C_i\widehat{\mb{x}}_i(k)|$ (stateless)  or the history of its residuals (stateful). Probabilistic thresholds can be designed based on the statistics of the noise terms for both cases. For the stateless case, define the threshold for a given  fault-alarm-rate (FAR) $\varkappa$ (see details in \cite[Fig.~3]{tnse21}) as,
\begin{align} \label{eq_thresh}
    \mc{T}_\varkappa = \sqrt{2}\mbox{erf}^{-1}(\varkappa)\Phi
\end{align}
with $\mbox{erf}^{-1}(\cdot)$ denoting the inverse Gauss error function and $\Phi$ is defined (based on $\|\mathbb{E}(\eta_k \eta^\top_k)\|_2$) approximately equal to \cite{khan2014collaborative,TCNS2020}, 
\begin{align} \label{eq_phi}
    \Phi :\simeq \|R\|_2 + \|H_i\|_2\frac{a_1N\|Q\|_2+a_2 \|\overline{R}\|_2}{N\overline{b}} 
\end{align}
where $\overline{R} := \mbox{diag}[H_i^\top R H_i]$, $a_1 := \|I_{Nn}- K D_C\|_2^2$, $a_2 := \|K\|_2^2$, and some $\overline{b}<1$. For the threshold $\mc{T}_\varkappa$, if the residual $r_i(k)>\mc{T}_\varkappa$ the sensor raises the alarm and declares fault (with FAR less than $\varkappa$). For the stateful case, first define the \textit{distance measure} based on the residual history over a \textit{sliding time-window} $\theta$ as,
    \begin{align} \label{eq_z}
    \mb{z}_i(k) = \sum_{m=k-\theta+1}^k \frac{(\mb{r}_{i}(k))^2}{\Phi},~k \geq \theta 
\end{align}  
It can be shown that $\mb{z}_i(k)$ follows the $\chi^2$-distribution  \cite{umsonst2019tuning}. Then, the threshold can be designed as \cite{icas21_attack},
\begin{align} \label{eq_thresh2}
    \overline{\mc{T}}_\varkappa = 2\Gamma^{-1}(1-\varkappa,\frac{\theta}{2})
\end{align}
with $\Gamma^{-1}(\cdot,\cdot)$ denoting the \textit{inverse regularized lower incomplete gamma function}. 
One can further design $\kappa$-redundant distributed estimators using the notion of $\kappa$-connected graph topology. Recall that a graph (network) is $\kappa$-edge-connected if it remains SC after removal of up-to $\kappa$ links (i.e., any subset of links of size $\kappa$); similarly, it is $\kappa$-node-connected if it remains SC after removal of up-to $\kappa$ (sensor) nodes (i.e., any subset of nodes of size $\kappa$). This is also referred to as \textit{survivable network design} \cite{umsonst2019tuning}. Note that this design procedure results from our distributed observability condition (instead of local observability), which only requires the strong connectivity of the network. To design a $\kappa$-connected sensor network, for example, the efficient algorithms in \cite{lau2009survivable,jabal2021approximation} can be used. Further, in the case of undirected networks (mutual linking among sensors), structural cost-optimal design of the sensor network can be considered to ensure a spanning tree (minimal connectivity) while optimizing the communication costs, e.g., see \cite{tnse19};
however, such a design problem is in general NP-hard for \textit{directed} SC networks. Algorithm~\ref{alg_0} summarizes our proposed methodology for distributed estimation (tracking), detection, and fault-tolerant design. To give an overview, the algorithm designs the network topology to fulfill (redundant) distributed observability, then uses TDOA measurements for distributed estimation; and finally, applies local detection techniques at every sensor.

\begin{algorithm} [t] \label{alg_0}
	\textbf{Input:} matrices $F,G$ (or sampling period $T$), FAR $\varkappa$, redundancy factor $\kappa$ \\
	Design the networks $\mc{G}$ (the sensor network) to be SC and $\kappa$-connected, e.g., via algorithms in \cite{lau2009survivable,jabal2021approximation}\;
	Calculate the TDOA measurement matrix $H_i$ via \eqref{eq_tdoa_new} based on the relative positions of sensors\;
	Design the block-diagonal gain matrix $K$ via LMI in \cite[Appendix]{jstsp}\;
	\Begin(sensor $i=1:N$ at every time $k$){Receives estimates $\widehat{\mb{x}}^j_{k-1|k-1}$ and TOAs  $\|\mb{p}(k)-\mb{p}_j(k)\|=ct_j$ from sensors $\{j|i \in \mc{N}(j)\}$\;
	Finds the TDOA measurements via \eqref{eq_h_simp}\;
	Finds the estimate $\widehat{\mb{x}}^i_{k|k}$ via Eq.~\eqref{eq_p}-\eqref{eq_m} on the target position (state) via the received information\;
	Finds the residual $r_k^i$  or distance measure $z_k^i$ \;
	Finds the threshold $\mc{T}_\varkappa$ or $\overline{\mc{T}}_\varkappa$ for the FAR $\varkappa$\;
	\If{$r_{k}^i> \mc{T}_\varkappa$ or $z_{k}^i> \overline{\mc{T}}_\varkappa$}{
		Sensor $i$ raises the fault alarm (detection probability $1-\varkappa$)\;
	    Sensor $i$ is isolated to avoid spread of the faulty measurement over the network\;	The number of faulty sensors $N_f=N_f+1$\;
	    \If{$N_f\leq \kappa$}{Distributed observability holds for tracking\;}\Else{Distributed tracking is terminated\;}
	}}
	\textbf{Output:} estimate of the target state $\widehat{\mb{x}}^i_{k|k}$, fault decision at sensor $i$, number of faulty sensors $N_f$	
	\caption{distributed $\kappa$-redundant estimation and localized FDI.}
\end{algorithm}

\section{Simulation} \label{sec_sim}
\begin{figure*} [t]
		\centering
 		\includegraphics[width=1.7in]{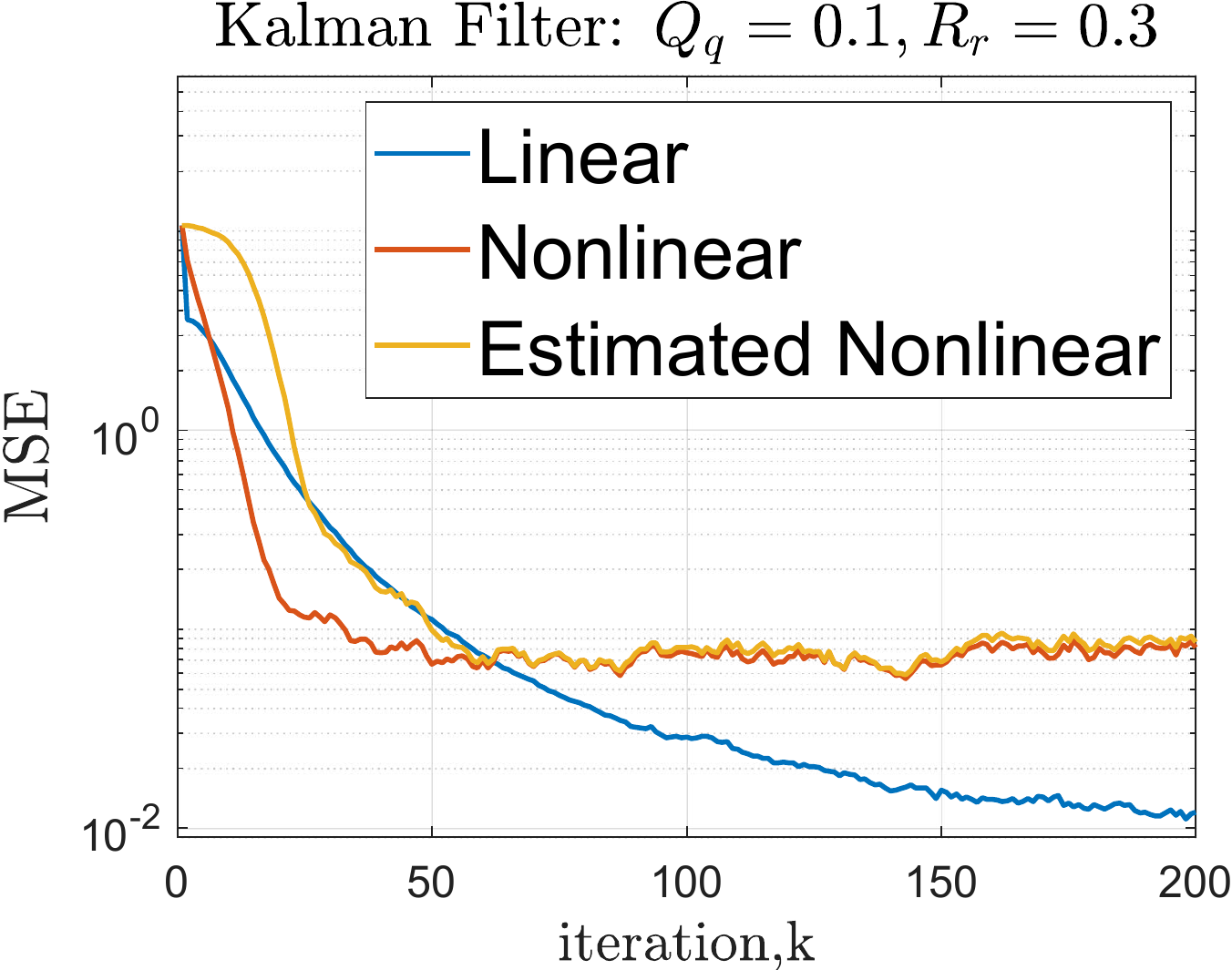}
        \includegraphics[width=1.7in]{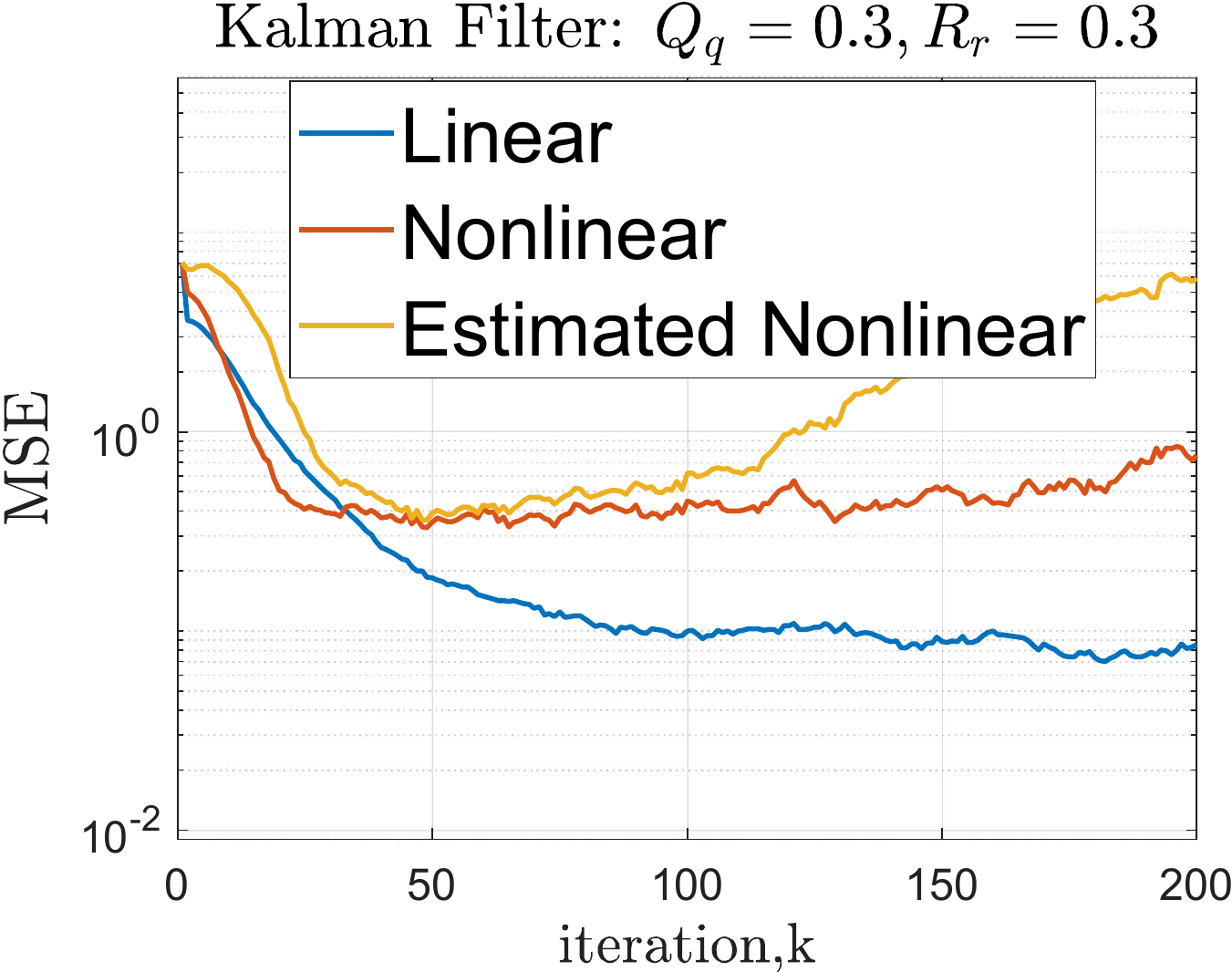}
 		\includegraphics[width=1.7in]{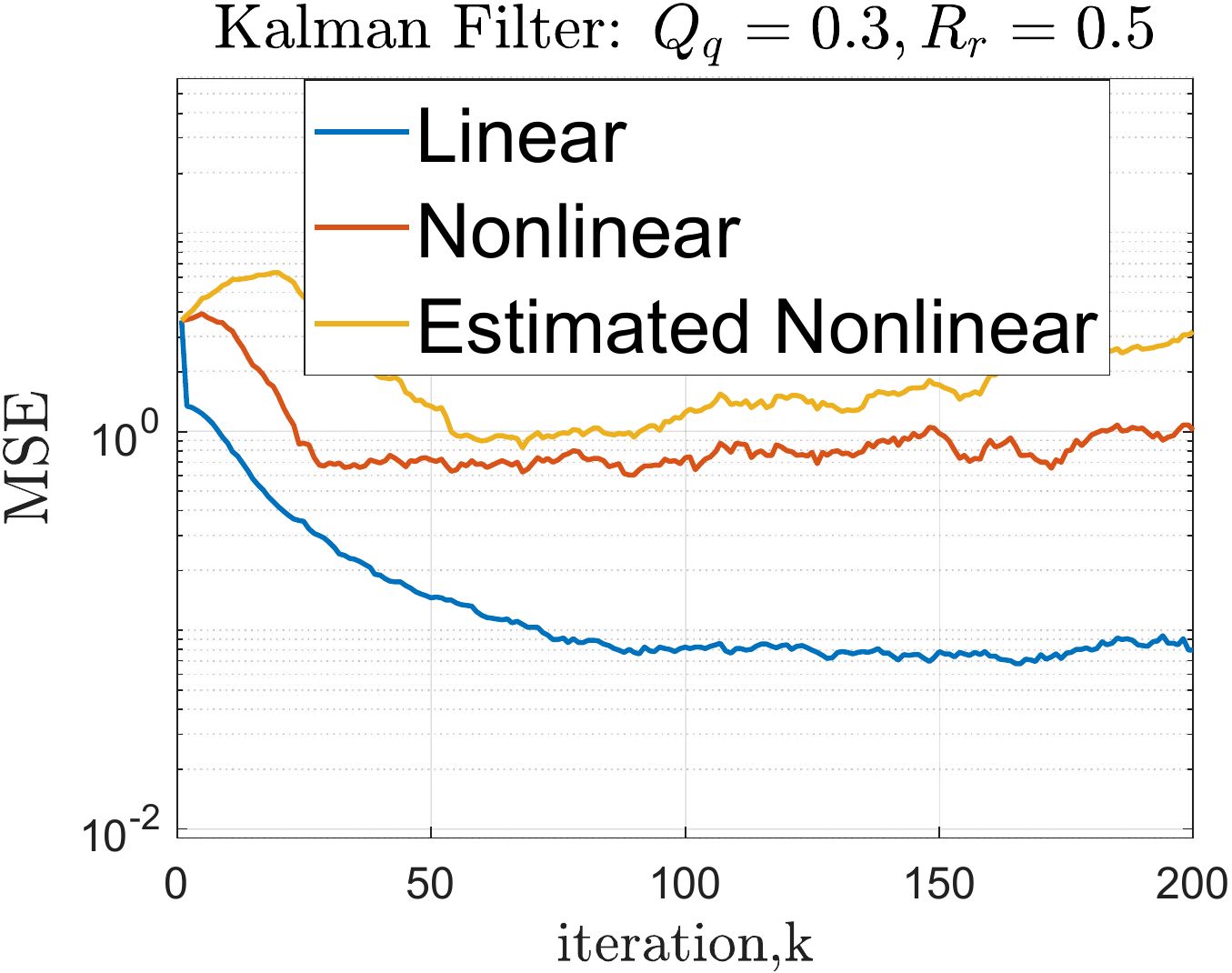}
        \includegraphics[width=1.7in]{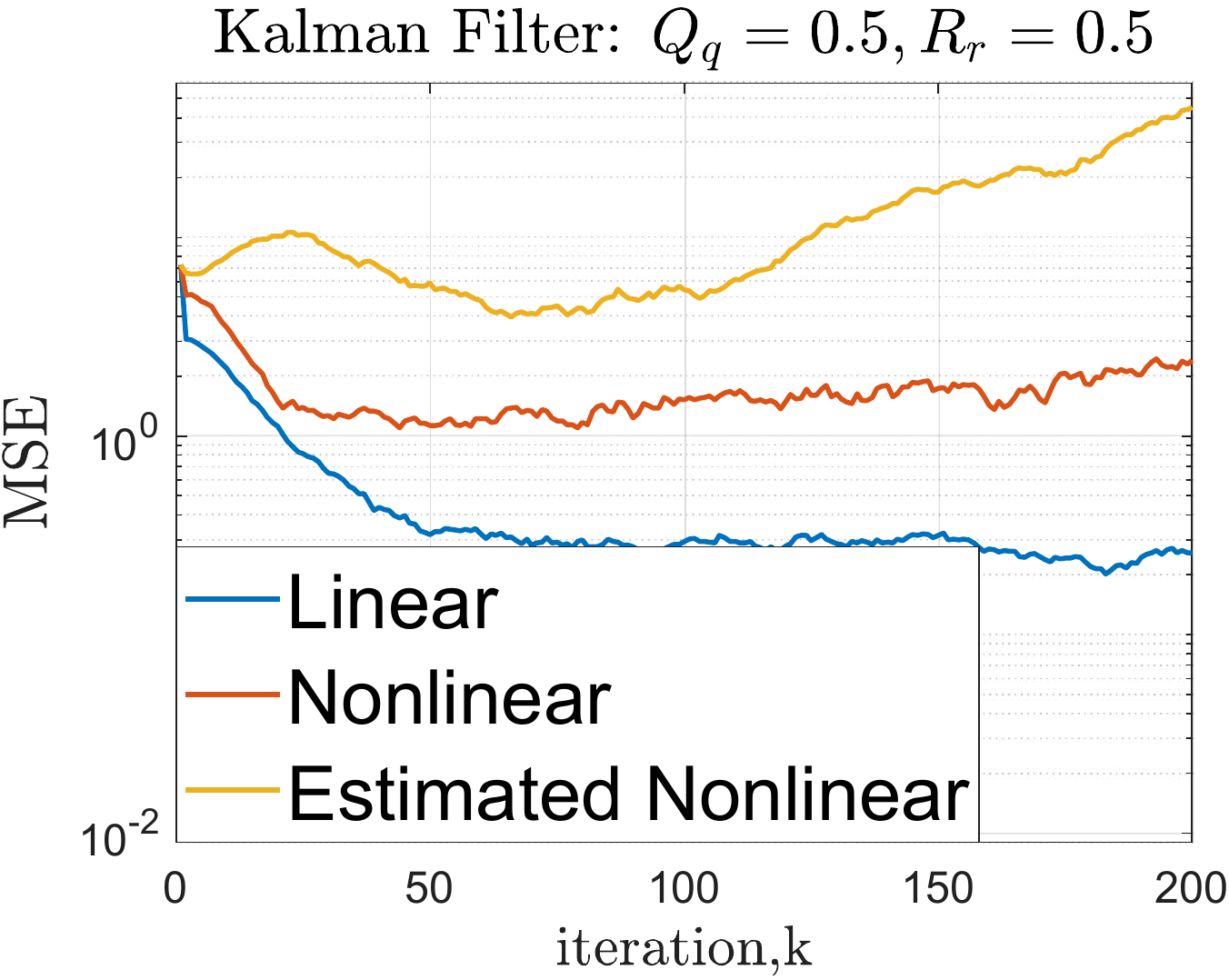}        
		\caption{This figure compares the MSE performance of the Kalman filter using the linear measurement model \eqref{eq_h_simp}-\eqref{eq_tdoa_new} (blue) versus the linearized (nonlinear) model \eqref{eq_h_ncv}-\eqref{eq_hij} with different (process and measurement) noise variances. %(TopLeft) $Q_q = 0.1,R_r=0.3$,  (TopRight) $Q_q = 0.3,R_r=0.3$, (BottomLeft) $Q_q = 0.3,R_r=0.5$, and (BottomRight) $Q_q = 0.5,R_r=0.5$. 
		For the nonlinear case, both exact and estimated position of the target are considered in  \eqref{eq_h_ncv}-\eqref{eq_hij}. Clearly, using the estimated target position in the output matrix $H_i$ degrades the MSE performance (yellow versus red).}
		\label{fig_kf}
	\end{figure*}

For the simulation, first, we compare the mean-square-error (MSE) performance of the proposed linear measurement model \eqref{eq_h_simp}-\eqref{eq_tdoa_new}  with the linearized (nonlinear) model \eqref{eq_h_ncv}-\eqref{eq_hij} using the Kalman Filter. We consider a sensor network of $N=10$ sensors over an all-to-all communication network (the centralized tracking setup) for  measurement and process noise $\mb{w}(k)=\mc{N}(0,Q_q^2\mb{I}_{6})$ and  $\nu=\mc{N}(0,R_r^2\mb{I}_{N-1})$ in \eqref{F_ncv} (with $T=0.1$). For the linearized model in \cite{ennasr2016distributed,ennasr2020time}, we consider two cases: (i) using the exact target position $\mb{p}$, and (ii) using the estimated target position $\widehat{\mb{p}}$ (which is used for the gain design). Note that the latter is only for the sake of comparison in terms of gain design accuracy in the distributed case. We consider the  coordinates of the initial locations of the sensors and the target randomly in the range $[0,10]$ and run the filter over $100$ Monte-Carlo trials. The results are shown in Fig.~\ref{fig_kf}.  
\begin{figure*} [h]
		\centering
 		\includegraphics[width=2.3in]{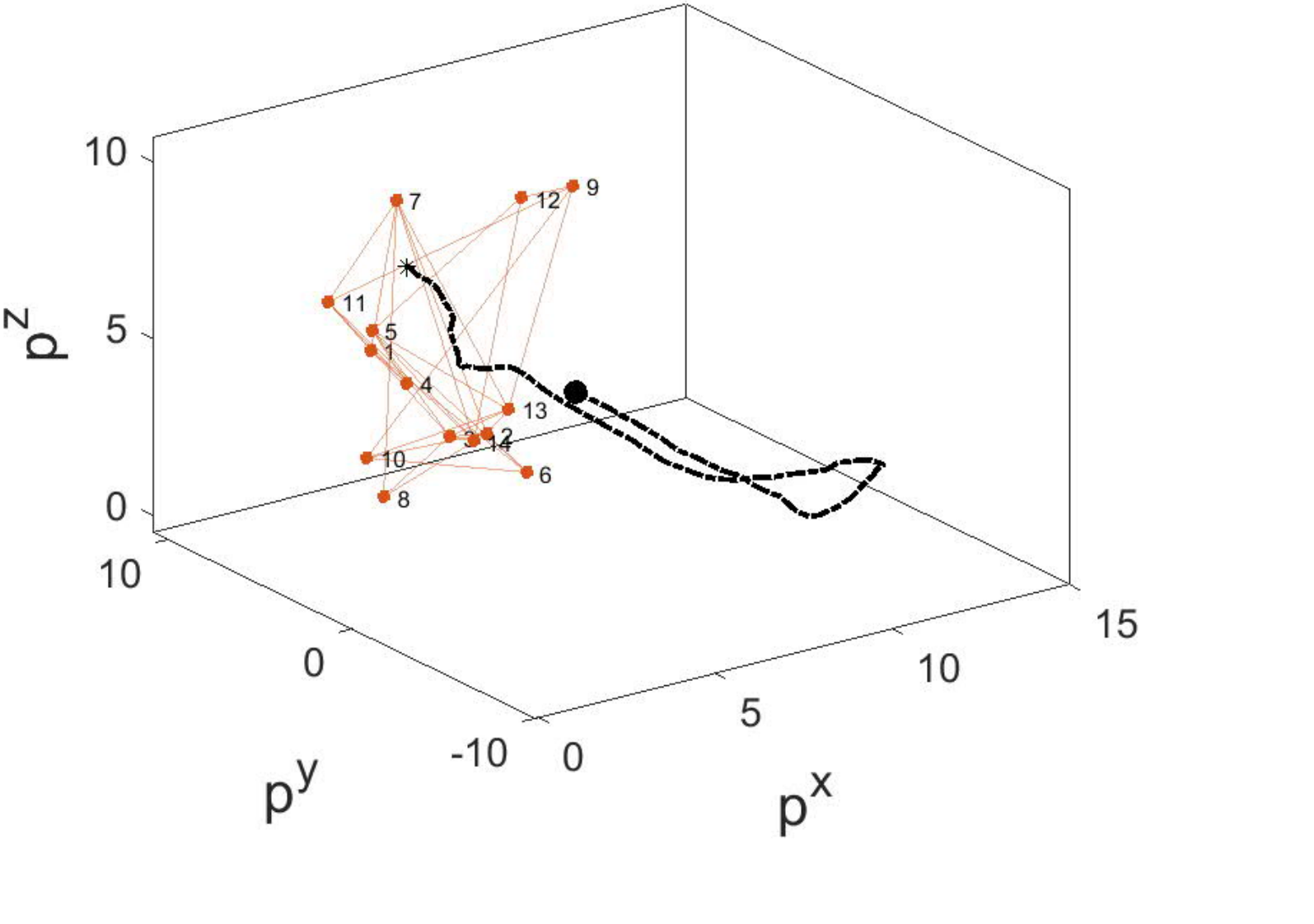}
        \includegraphics[width=2.3in]{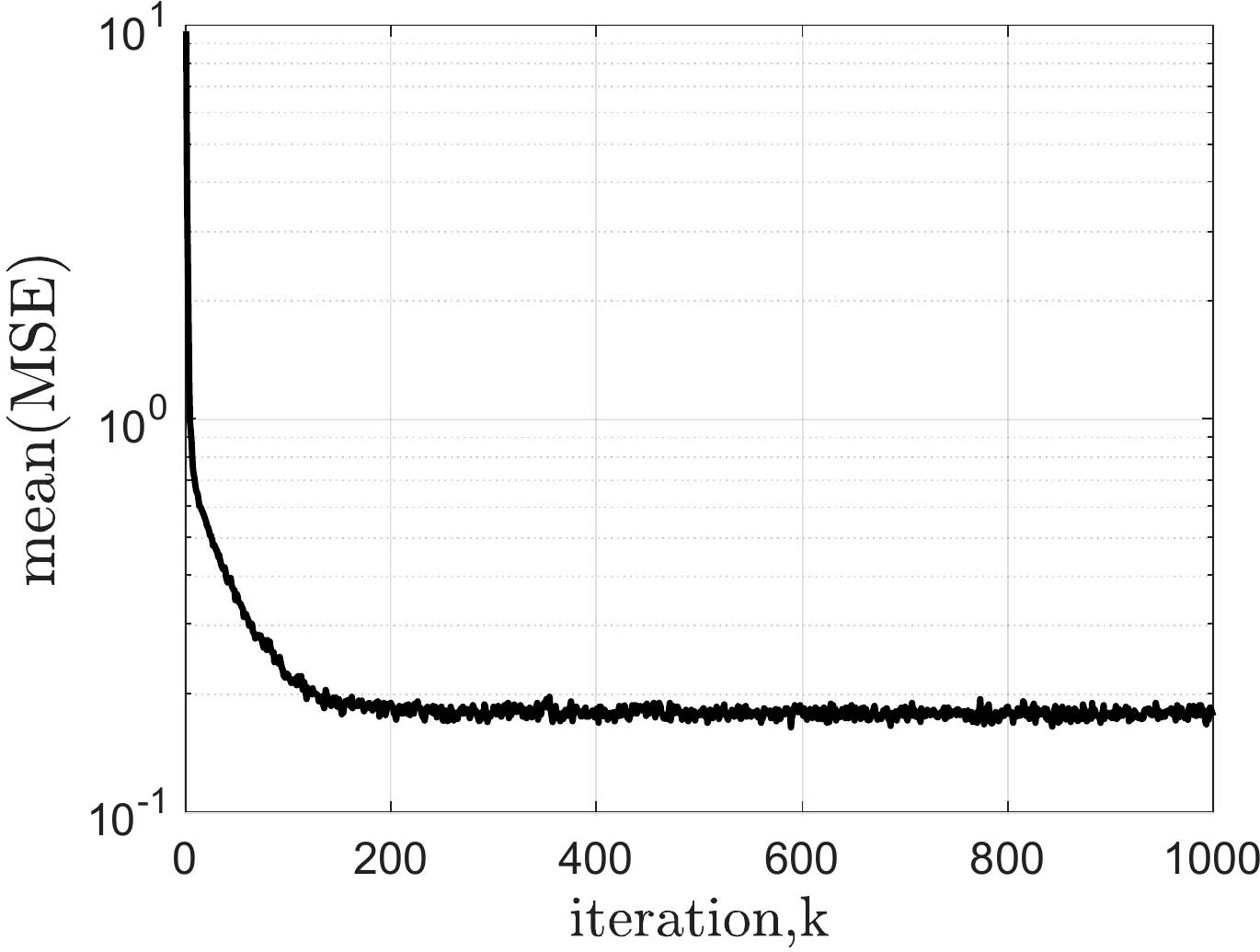}
 		\includegraphics[width=2.3in]{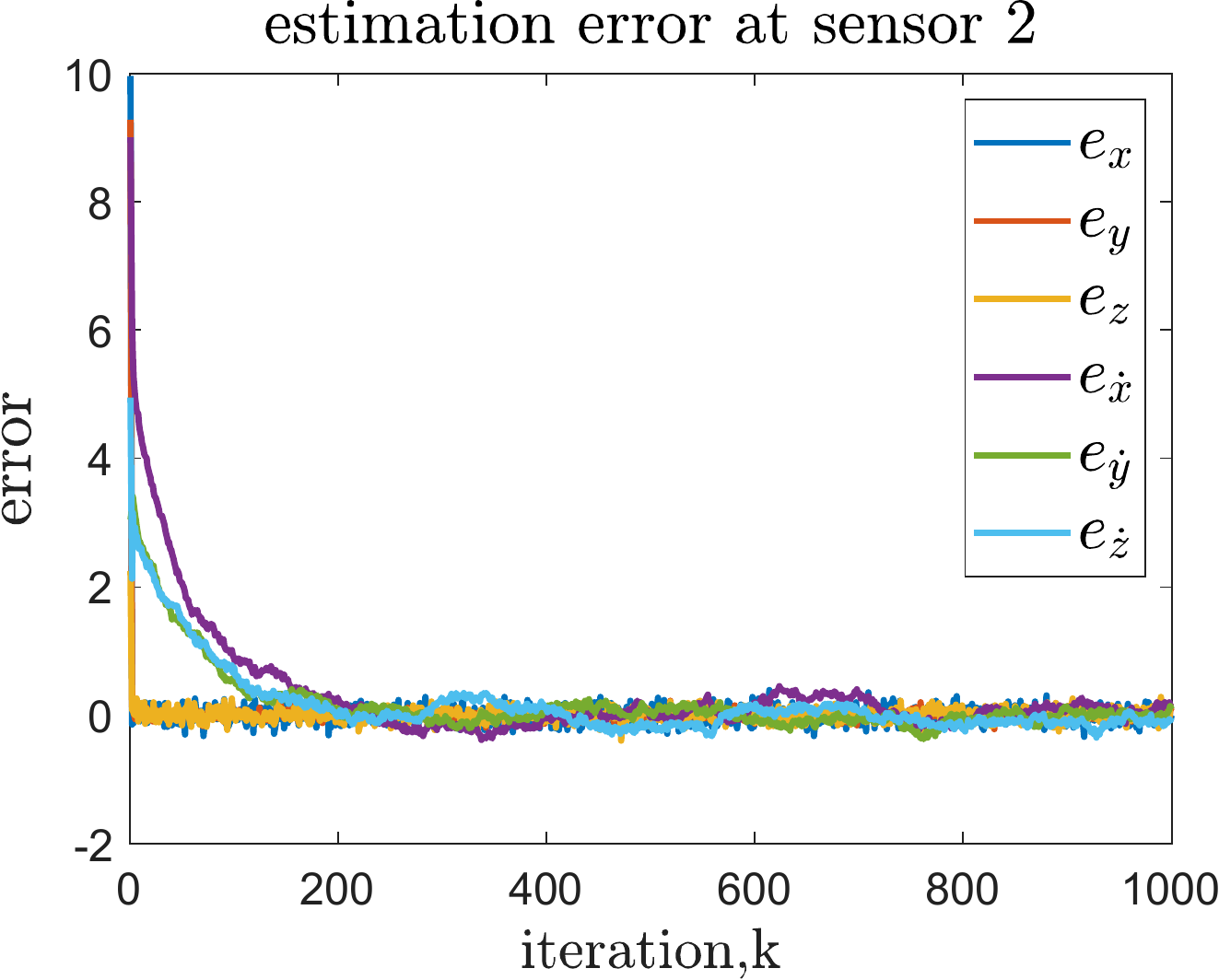}
        \includegraphics[width=2.3in]{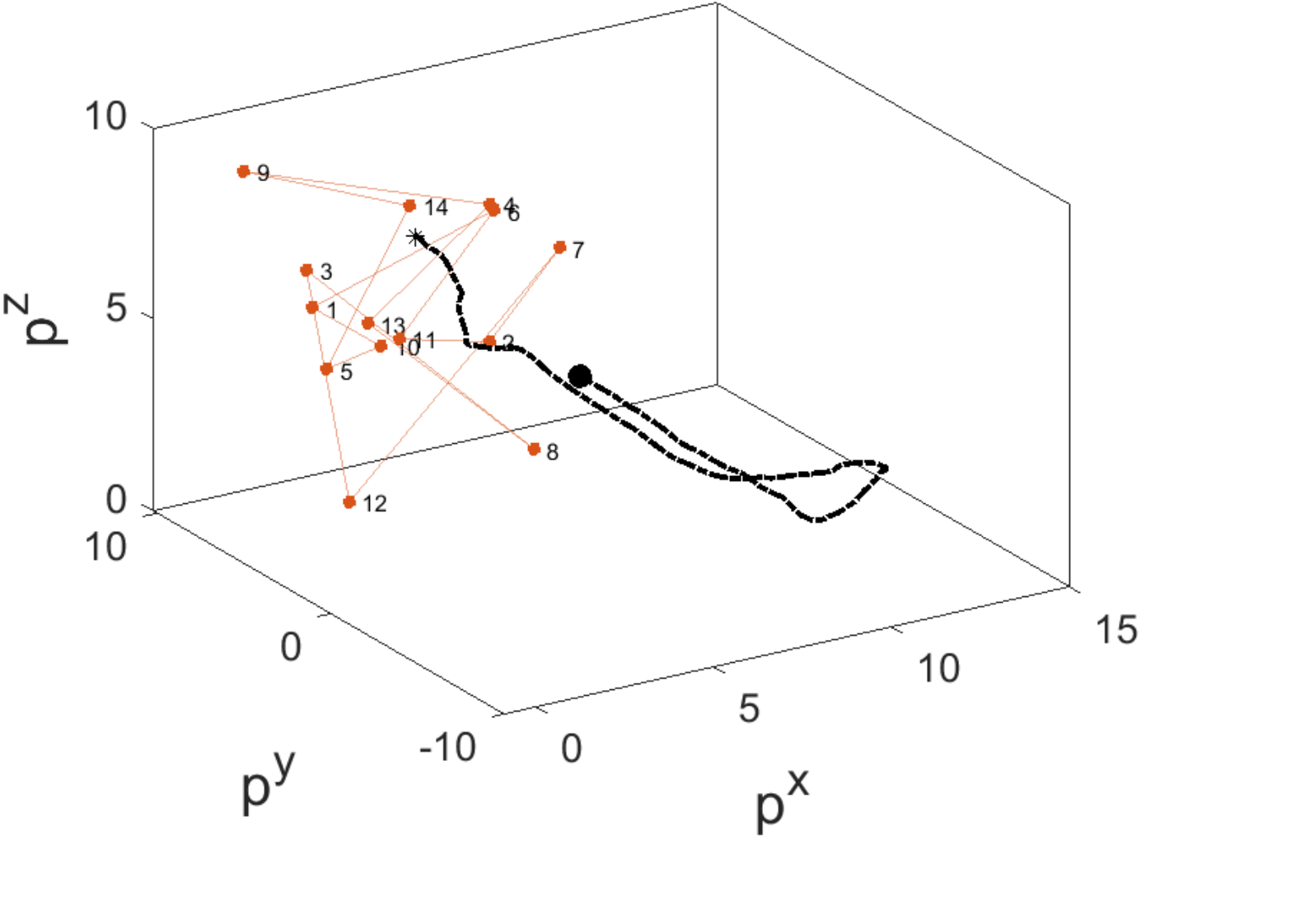} 
        \includegraphics[width=2.3in]{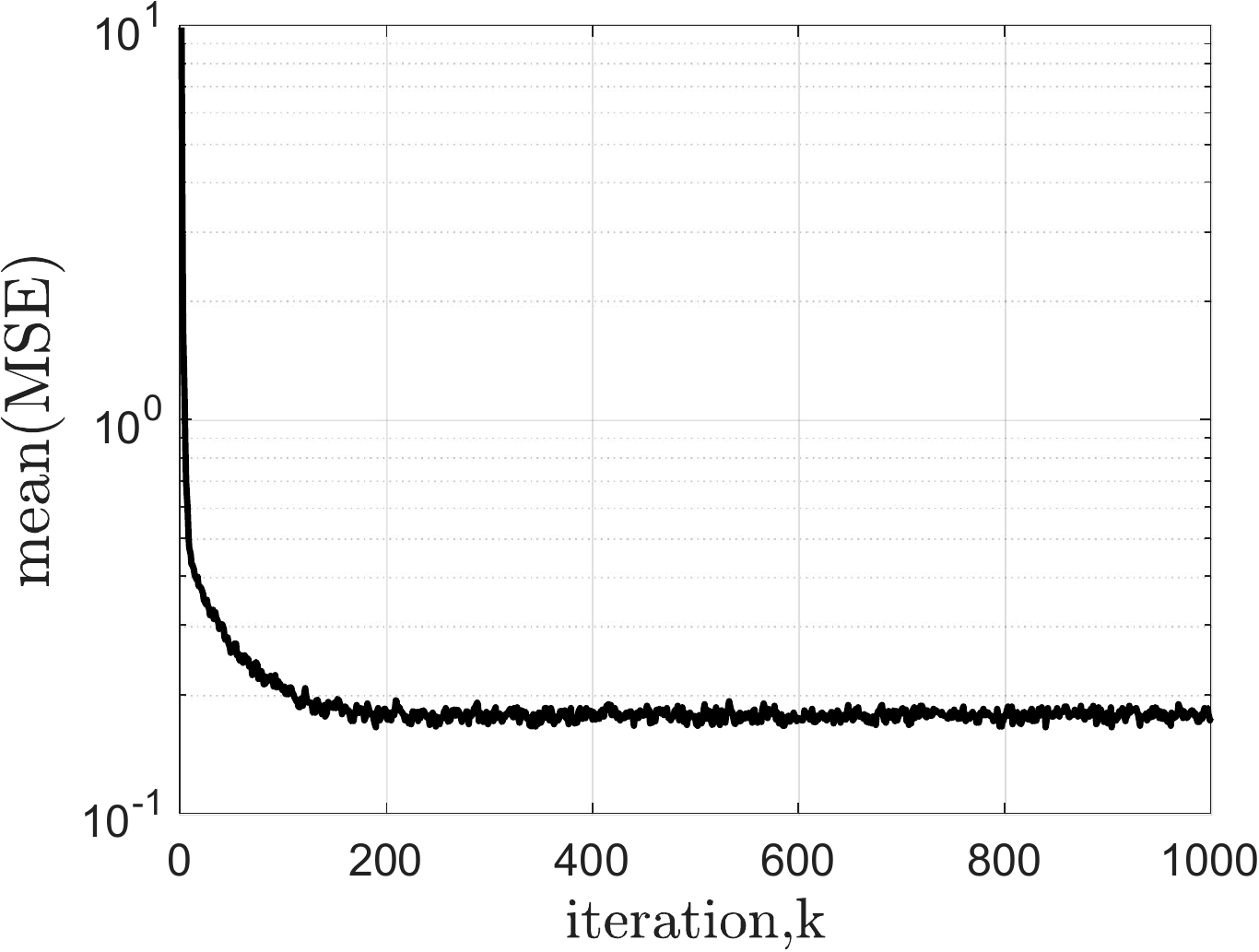}
 		\includegraphics[width=2.3in]{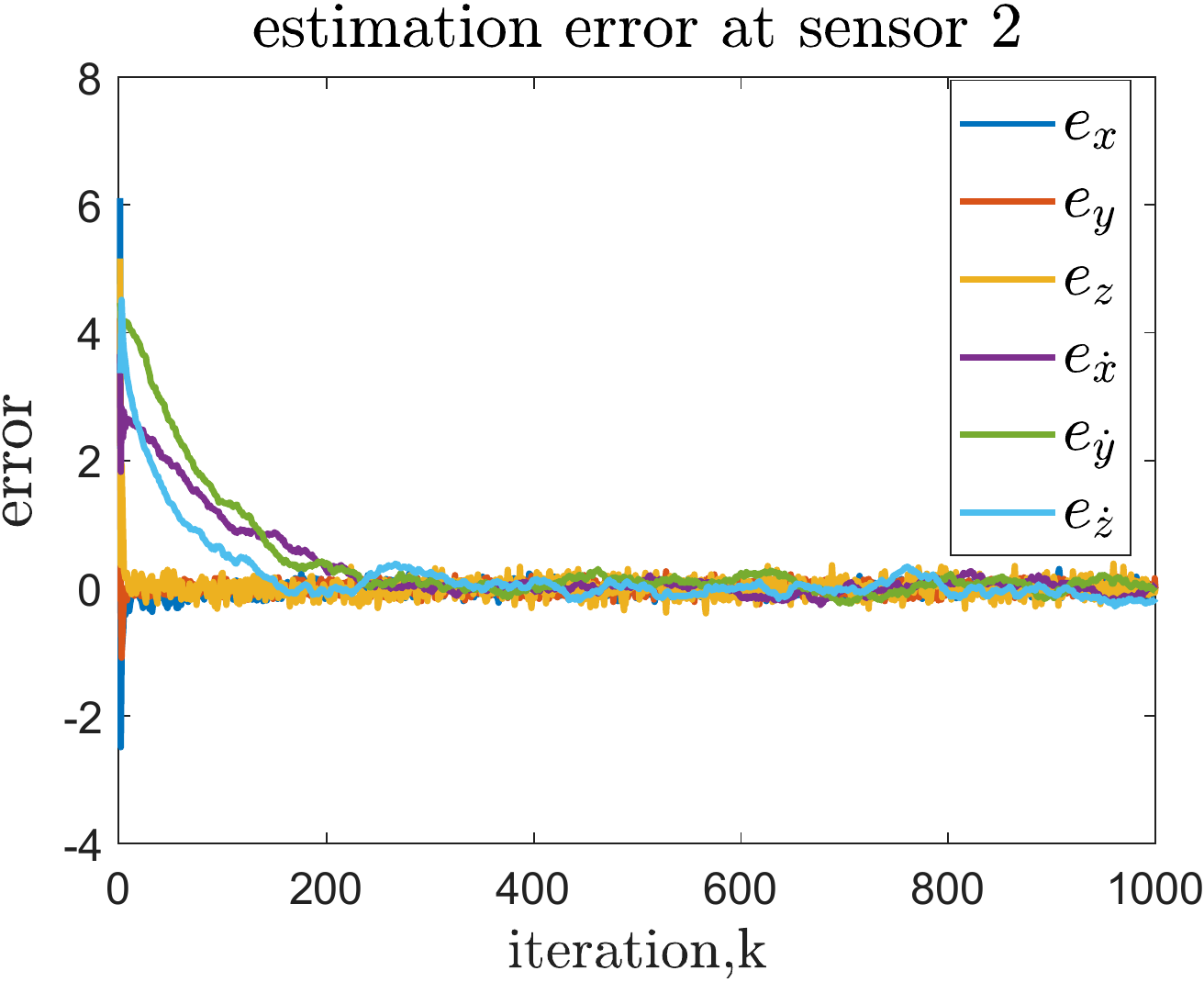}      
		\caption{This figure shows the Monte-Carlo performance of the proposed distributed STS protocol \eqref{eq_p}-\eqref{eq_m}: (Top) over a $2$-connected network, (Bottom) over a cyclic network with reduced connectivity (after link removal) and new set of sensor positions. Clearly, the estimation errors on the target velocity converge slower than the  errors on the position (right figures). The target path is shown by a black dashed line with final position as the big black circle in the left figures with the sensor network shown in red.}
		\label{fig_dist} 
\end{figure*}	
As it is clear from the figure, the proposed method considerably improves the MSE of the Kalman filter for higher process and measurement noise. Further, as discussed in Remark~\ref{rem_compare}, in the distributed linearized setups the estimated position of the target is used in the output matrix $H_i$ for practical gain design, which further degrades the error stability performance due to additive uncertainty on $H_i$s. 

On the same NCV setup, we provide MSE performance analysis of the proposed distributed STS estimator \eqref{eq_p}-\eqref{eq_m} in Fig.~\ref{fig_dist}. 
The estimation error is bounded steady-state stable (averaged at all sensors) and unbiased in steady-state (the state errors at sensor $2$ are shown as an example). The simulation is performed once over $\kappa=2$-connected network and also over a simple cyclic network. Distributed observability is preserved after reducing the network connectivity to the cyclic network (which is SC) after removing (or missing) some (faulty) communication links (a survivable network design). The consensus fusion weights are random while satisfying stochastic property for consensus \cite{rikos2014distributed}. Note that, for each case, the local feedback gain is designed only once (using the LMI in \cite[Appendix]{jstsp}) since the fusion weight matrix $W$ and the measurement matrix $H_i$ (given by \eqref{eq_h_simp}-\eqref{eq_tdoa_new}) are constant. This scenario is computationally more efficient (and more accurate) as compared to the repetitive gain design in \cite{ennasr2020time,ennasr2016distributed} which is performed at every iteration due to time-varying linearized $H_i$ in \eqref{eq_h_ncv}-\eqref{eq_hij} (which is approximated by the estimated target position). 

%Recall that this estimator performs $L$ additional steps of consensus and communications between every two time-steps $k$ and $k+1$ of the system dynamics (see Table~\ref{tab_compare}). This increases both the computational load and communication traffic over the network, while may improve the MSE performance due to more information exchange over the sensor network. For the simulation, we consider a random  network of $N=14$ spatially distributed sensors. The network is undirected (mandated by \cite{he2020secure}) with symmetric stochastic weights. 

\section{Conclusion and Future Directions} \label{sec_con}
This paper provides a distributed (single-target) tracking with less communication and computational requirement than the existing (STS and DTS) methods. In some applications, one can extend the results to consider a rigid formation of leader-follower UAVs (and smart mobile sensors) \cite{7995044,olfati2002distributed} to track a long-range maneuvering target (that may move out of the short-range of static sensors). Furthermore, the notions of optimality and  energy efficiency can be further addressed as future research directions to design minimal cost networks in terms of communication cost and energy consumption via the results in \cite{zheng2016energy,pequito2014optimal,tnse19,farjam2020power}. Addressing delay-tolerant protocols is another promising research direction \cite{abolfazli2021time}

\bibliographystyle{IEEEbib}
\bibliography{bibliography}

\end{document}